\documentstyle[12pt]{article}
\begin{document}
\begin{center}
{\large\bf Astrophysical thermonuclear functions for Boltzmann-Gibbs statistics and Tsallis statistics}\\[0.5cm]
R.K. Saxena\\ 
Department of Mathematics and Statistics, Jai Narain Vyas University, Jodhpur 342001, India\\[0.3cm]
A.M. Mathai\\ 
Department of Mathematics and Statistics, McGill University,\\ 
Montreal, Canada H3A 2K6 \\[0.3cm]
H.J. Haubold\\ 
Office for Outer Space Affairs, United Nations\\
P.O. Box 500 A-1400 Vienna, Austria \\[0.3cm]
\end{center}
{\bf Abstract}

We present an analytic proof of the integrals for astrophysical thermonuclear functions which are derived on the basis of Boltzmann-Gibbs statistical mechanics. Among the four different cases of astrophysical thermonuclear functions, those with a depleted high-energy tail and a cut-off at high energies find a natural interpretation in q-statistics.
\section{Introduction}

One of the first applications of Gamow's theory of quantum mechanical potential barrier penetration to other than his analysis of alpha radioactivity was in the field of thermonuclear astrophysics [1,2]. Atkinson and Houtermans proposed that the source of energy released by stars lay in thermonuclear reactions taking place near their centres where the motion of nuclei was supposed to be in thermal equilibrium. The state of hot stellar plasmas is such that only the lightest chemical elements could contribute because of the Coulomb repulsion between nuclei. In effect, the rate of energy production is goverend by the average of the Gamow penetration factor over the Maxwell-Boltzmann velocity distribution [3]. The thermonuclear reaction rate is the coefficient in the rate equation that is used to describe the change of chemical composition of hot plasmas. Similarly fundamental, the inverse of this coefficient is the characteristic time scale for the respective thermonuclear reaction. Nuclear and neutrino astrophysics are vibrant fields with many experiments collecting data continuing exploring energy production in stars, particularly the Sun, deduced from neutrino measurments [4].

Commonly, hot stellar fusion plasmas are described in terms of Boltzmann-Gibbs statistical mechanics based on the entropy
$S_{BG}=-k\sum^W_{i=1}p_i lnp_i$.
Thermodynamical and ststistical description of nonextensive systems may require a generalization of Boltzmann-Gibbs thermostatistics. Examples of physical systems or processes where Boltzmann-Gibbs approach seems to be inadequate are large self-gravitating systems and hot and turbulent plasmas.

An ultimate generalization of Boltzmann-Gibbs thermostatistics is due to Tsallis [5] based on the following expression for entropy
\begin{equation}
S_q=\frac{1-\sum^W_{i=1}p_i^q}{q-1},\;\;q \in R, S_1=S_{BG}, \;\;\sum^W_{i=1}p_i=1,
\end{equation}
with the following expression for equal probabilities
\begin{equation}
S_q(p_i=1/W, \forall_i)=k\frac{W^{1-q}-1}{1-q}\equiv kln_qW,
\end{equation}
$$S_1(p_i=1/W, \forall_i)=klnW=S_{BG}.$$

The optimization of the entropic form (1) under the restriction
\begin{equation}
<H>_q\equiv\frac{\sum^W_{i=1}p_i^qE_i}{\sum^W_{j=1}p_j^q}=U_q,
\end{equation}
where $<\ldots>_1=<\ldots> p_i^q/\sum^W_{j=1}p_j^q$
is the escort distribution, $\{E_i\}$
are the eigenvalues of the Hamiltonian $H$
with appropriate boundary conditions, and the internal energy $U$
is a finite fixed value, yields
\begin{equation}
p_i=\frac{e_q^{-\beta_q(E_i-U_q)}}{\bar{Z}_q},
\end{equation}
with
$$\bar{Z}_q\equiv\sum^W_{j=1}e_q^{-\beta_q(E_j-U_q)},\;\; \beta_q\equiv\frac{\beta}{\sum^W_{j=1}p_j^q},$$
$\beta=1/kT$. Tsallis [5] verifies that $q=1$
recovers Boltzmann-Gibbs weight, $q>1$
implies a power-law tail at high values of $E_i$, and  $q<1$
implies a cut-off at high values of $E_i$.
Subsequently, this behavior is reflected in the equilibrium distribution of energies.
In the following, a method for obtaining closed-form representations of thermonuclear reaction rates (astrophysical thermonuclear functions) is developed, taking into account the above described behavior of q-statistics, these are the cases $q>1$ (dubbed ``depleted'') and $q<1$ 
(dubbed ``cut-off'') for the energy distribution. Thus, q-statistics provides a natural physical interpretation for hot plasmas where deviations from the Maxwell-Boltzmann distribution are expected.

\section{Definition of astrophysical thermonuclear functions $I^d(\nu-1, a,z,\rho)$}

By employing  a statistical technique, Haubold and Mathai [6] have established the following integral formula for the derivation of closed-form representations for four astrophysical thermonuclear functions:
\begin{eqnarray}
\int^d_0&y^{\nu-1}&exp[-ay-zy^{-\rho}]dy \stackrel{def}{=}I^d(\nu-1,a,z,\rho)\\\nonumber  
&=&\frac{d^\nu}{\rho}\sum^\infty_{r=0}\frac{(-ad)^r}{(r)!}H^{2,0}_{1,2}\left[\frac{z^{1/\rho}}{d}|^{(\nu+r+1,1)}_{(0,1/\rho)(\nu+r,1)}\right](\rho\neq 0), for\;\; d<\infty;\\\nonumber
&=&\frac{a^{-\nu}}{\rho}H^{2,0}_{0,2}\left[az^{1/\rho}|_{(0,1/\rho),(\nu,1)}\right]\,\,(\rho\neq 0), \mbox{for}\,\, d= \infty, 
\end{eqnarray}
where $Re(\nu)>0, Re(a)>0, Re(z)>0, Re(\rho)>0.$ 
         It has been shown that by the application of the Mellin-Barnes integral rerpresentation of the exponential function, the given integral can be transformed into a Mellin-Barnes type integral representing an H-function [7]. We also deduce the value of this integral when $\rho<0$.   Results for all the four astrophysical thermonuclear functions are also obtained. The integral $I_3(z,t,\nu)$ [8, see also 2,9] is evaluated in a generalized form.  For the sake of simplicity and continuation we adopt the same notations as used in  [6, see also 2,9].
        In order to prove the integral (5), we employ the following Mellin-Barnes integral representation for the exponential function, namely [10,11] 
\begin{equation}
exp[-x]=\frac{1}{2\pi\omega}\int_L\Gamma(s)x^{-s}ds, |x|<\infty.
\end{equation}                                       
where $L$ is a suitable contour $(\omega=(-1)^{\frac{1}{2}})$, interchange the order of integration,  the given integral, denoted by $I^d(\nu-1,a,z,\rho)$), transforms into the form
\begin{equation}
I^d(\nu-1,a,z,\rho)=\frac{1}{2\pi\omega}\int_L\Gamma(s)z^{-s}\int_0^dy^{\nu+\rho s-1} exp(-ay)dy ds
\end{equation}   
We know that 
\begin{eqnarray}
\int^d_0& exp(-ay)&y^{\nu-1}dy\\\nonumber
&=&d^\nu\sum^\infty_{r=0}\frac{(-ad)^r\Gamma(\nu+r)}{\Gamma(\nu+r+1)(r)!} \mbox{for}\;\;d<\infty, Re(\nu)>0\\
\nonumber
&=&\Gamma(\nu)/a^\nu, \mbox{for}\,\,d=\infty\;\; Re(a)>0, Re(\nu)>0.
\end{eqnarray}                       
         By virtue of the above values of the y-integral for $d<\infty$ and for  $d=\infty$, it is found that 
\begin{eqnarray}
I^d(\nu-1, a,z,\rho) &=&\frac{d^\nu}{\rho}\sum^\infty_{r=0}\frac{(-ad)^r}{(r)!}\int_L\frac{\Gamma(s/\rho)\Gamma(\nu+s+r)z^{-s/\rho}}{\Gamma(\nu+s+r+1)}ds,\label{line1}\\
&(\rho\neq 0)&\mbox{for}\;\;d<\infty\nonumber\\ 
&=&\frac{a^{-\nu}}{2\pi\omega\rho}\int_L\Gamma(\frac{s}{\rho})\Gamma(\nu+s)a^{-s}z^{-s/\rho}ds,\nonumber\\
&(\rho\neq 0)& \mbox{for}\;\; d=\infty\nonumber
\end{eqnarray}

The integral formula  (5) now  readily follows from the equations (9), if we interpret the above contour integrals in terms of the H-function, which is  defined by means of a Mellin-Barnes type integral in the following manner  [7, p.2].
\begin{eqnarray}
H^{m,n}_{p,q}(z)&=&H^{m,n}_{p,q}\left[z\left|^{(a_p,A_p)}_{(b_q,B_q)}\right.\right]\label{line1}\\
&=&H^{m,n}_{p,q}\left[z\left|^{(a_1,A_1)\ldots(a_p,A_p)}_{(b_1,B_1)\ldots(b_q,B_q)}\right.\right]=\frac{1}{2\pi\omega}\int_L\Theta(\xi)z^{-\xi}d\xi\nonumber
\end{eqnarray}  
where
\begin{equation}
\Theta(\xi)=\frac{\left[\Pi^m_{j=1}\Gamma(b_j+B_j\xi)\right]\left[\Pi^n_{j=1}\Gamma(1-a_j-A_j\xi)\right]}{\left[\Pi^q_{j=m+1}\Gamma(1-b_j-B_j\xi)\right]\left[\Pi^\rho_{j=n+1}\Gamma(a_j+A_j\xi)\right]}
\end{equation}                           
an empty product is interpreted as unity ; $m, n, p, q\in N_0 \mbox{with}\;\;  0\leq n \leq p, 1\leq m\leq q,\\ 
A_i, B_j \in R_+, a_i, b_j \in R$ or $C( i=1,\ldots,p; j=1,\ldots,q)$, such that
\begin{equation}
A_i(b_j+k)\neq B_j(a_i-l-1)(k,l \in N-0; i=1,\ldots, n; j=1, \ldots, m).
\end{equation}                     
We employ the usual notations: $N_0=(0,1,2\ldots); \,\,R=(-\infty, \infty),R_+=(0,\infty)$ and C being the complex number field. $L$ is one of the contours described in the monograph by Mathai [11].        
In case we make the transformation $\rho=-\eta$ with $\eta>0$ in (9), then the following result holds.
\begin{eqnarray} 
I^d(\nu-1,a,z,-\eta)&=&-\frac{d^\nu}{\eta}\sum^\infty_{r=0}\frac{(ad)^r}{(r)!}H^{1,1}_{1,2}\left[z^{1/\eta}\left|^{(1-\nu-r,1)}_{(0,1/\eta),(-\nu-r,1)}\right.\right]\label{line1}\\
&\eta\neq 0,& \mbox{for}\;\; z<\infty;\nonumber\\      
&=&-\frac{a^{-\nu}}{\eta}H^{1,1}_{1,1}\left[\frac{z^{1/\eta}}{a}\left|^{(1-\nu,1)}_{(0,1/\eta)}\right.\right]\nonumber\\
&\eta\neq 0,& \mbox{for}\;\;z=\infty,
\end{eqnarray}                                
where $Re(\nu)>0, Re(a)>0, \eta>0, Re(z)>0.$
When $\rho$ is real and rational then the H-functions appearing in (5), (13) and (14) can be reduced to G-functions by the application of the  well-known  multiplication formula for gamma functions [12]:
\begin{equation}
\Gamma(mz)=(2\pi)^{\frac{1-m}{2}}m^{mz-\frac{1}{2}}\Gamma(z)\Gamma(z+\frac{1}{m})\ldots\Gamma(z+\frac{m-1}{m}),\,\,m=1,2,\ldots
\end{equation} 
\section{The astrophysical thermonuclear functions $I_1(z,\nu)$ and $I_2(z,d,\nu)$}

As an application of the results (5), (13), and (14) we deduce the values of the following two thermonuclear functions $I_1$ [8] and $I_2$ [''cut-off'', 8,13,14,17], in terms of which the astrophysical thermonuclear functions are expressed [8, see also 2,9]:
\begin{eqnarray}
I_1(z,\nu) &\stackrel{def}{=}&\int^\infty_0y^{\nu-1}exp[-y-\frac{z}{y^{1/2}}]dy\\\nonumber 
&=& \pi^{-1/2}G^{3,0}_{0,3}\left[\frac{z^2}{4}\left|0,\frac{1}{2}, \nu\right.\right],
\end{eqnarray}
where $Re(z)>0, Re(\nu)\geq 0$ and $G^{3,0}_{0,3}(.)$is the Meijer's G-function [10].
\begin{eqnarray}
I_2(z,d,\nu)&\stackrel{def}{=}&\int^d_0 y^{\nu-1}exp[-y-zy^{-1/2}]dy\nonumber\\
&=& \frac{d^\nu}{\pi^{1/2}}\sum^\infty_{r=0}\frac{(-d)^r}{(r)!}G^{3,0}_{1,3}\left[\frac{z^2}{4d}\left|^{(\nu+r+1)}_{(\nu+r), 0,1/2}\right.\right];
\end{eqnarray}                                 
where $d>0, Re(z)>0, Re(\nu)\geq 1$.

\section{The astrophysical thermonuclear functions $I_3(z,t,\nu,\mu)$ and $I_4(z,\delta,b,\nu)$}

We now show that [8]
\begin{eqnarray}
I_3(z,t,\nu,\mu)&\stackrel{def}{=}&\int^\infty_0 y^{\nu-1}exp[-\left\{y+z(y+t)^{-\mu}\right\}]dy\label{line1}\\  
&=&t^\nu\sum^\infty_{r=0}\frac{\Gamma(\nu+r)}{(r)!}t^r H^{2,1}_{2,3}\left[zt^{-\mu}\left|^{(1+\nu,\mu),(\nu+r+1,\mu)}_{(0,1),(1+\nu,\mu),(1,\mu)}\right.\right]\nonumber\\
&+&\sum^\infty_{r=0}\frac{t^r}{(r)!}H^{2,2}_{2,4}\left[z\left|^{(1-r,\mu), (\nu,\mu)}_{(0,1),(\nu,\mu),(1,\mu)(\nu-r,\mu)}\right.\right],\nonumber
\end{eqnarray}                                     
where  $Re(\nu)> 0, Re(z)>0 \mbox{and}\;\; \mu>0$.
To prove (18) we observe that in view of the formula (6), the value of the integral is equal to 
\begin{equation} 
\frac{1}{2\pi\omega}\int_L\Gamma(s)z^{-s}\int^\infty_0y^{\nu-1}[exp(-y)](y+t)^{s\mu}dyds
\end{equation}           
Evaluating the $y$-integral in terms of the Whittaker function, defined in [12, p.255] and 
[12, p.257]:
\begin{eqnarray}
\Psi(a,c;z)&=&\frac{1}{\Gamma(a)}\int^\infty_0 x^{a-1}e^{-zx}(1+x)^{c-a-1}dx, Re(a)>0, Re(z)>0.\label{line1}\\
&=&\frac{\Gamma(1-c)}{\Gamma(a-c+1)}\Phi(a,c,z)+\frac{\Gamma(c-1)}{\Gamma(a)}z^{1-c}\Phi(a-c+1,2-c;z)\nonumber\\
&=&\frac{\Gamma(1-c)}{\Gamma(a-c+1)}\sum^\infty_{r=0}\frac{(a)_rz^r}{(c)_r(r)!}+\frac{\Gamma(c-1)}{\Gamma(a)}z^{1-c}\sum^\infty_{r=0}\frac{(a-c+1)_rz^r}{(2-c)_r(r)!},\nonumber
\end{eqnarray}
where $c$ is not an integer and $\Phi(a,c;z)$ is Kummer's confluent hypergeometric function [12,p.248], the integral  (19) reduces to an elegant formula  
\begin{equation}
\frac{\Gamma(\nu)t^\nu}{2\pi\omega}\int_L\Gamma(s)\Psi(\nu,\nu+s\mu+1;t)z^{-s}t^{s\mu}ds.
\end{equation}

 The integral (21) can be evaluated by expressing the Whittaker function appearing in its integrand  in terms of its equivalent series given by (20) above, and  reversing the order of integration and summation. Thus the given integral finally transforms into the form:
\begin{eqnarray}
&I_3&(z,t,\nu,\mu)=t^\nu\sum^\infty_{r=0}\frac{\Gamma(\nu+r)}{(r)!}\frac{t^r}{2\pi\omega}\int_L\frac{\Gamma(s)\Gamma(-\nu-s\mu)\Gamma(\nu+s\mu+1)}{\Gamma(-s\mu)\Gamma(\nu+r+s\mu+1)}z^{-s}t^{s\mu}ds\nonumber\\
&+&\sum^\infty_{r=0}\frac{t^r}{(r)!}\frac{1}{2\pi\omega}\int_L\frac{\Gamma(s)\Gamma(\nu+s\mu)\Gamma(r-s\mu)\Gamma(1-\nu-s\mu)}{\Gamma(-s\mu)\Gamma(1+r-\nu-s\mu)}z^{-s}ds,\label{line2}
\end{eqnarray}                                             
which on interpreting with the help of the definition of the H-function (10) establishes the desired result  (18). If we set $\mu=1/2$ in (18), then by an appeal to the duplication formula for the gamma function (this is equation (15) for $m=2$), the H-functions reduce to G-functions  [10]  and consequently, we obtain    
\begin{eqnarray}
I_3(z,t,\nu)&\stackrel{def}{=}&\frac{t^\nu}{\pi^{1/2}}\sum^\infty_{r=0}\frac{\Gamma(\nu+r)}{(r)!}t^rG^{3,1}_{2,4}\left[\frac{z^2}{4t}\left|^{(1+\nu),(\nu+r+1)}_{0,1/2,(1+\nu),1}\right.\right]\label{line1}\\
&+&\frac{1}{\pi^{1/2}}\sum^\infty_{r=0}\frac{t^r}{(r)!}G^{3,2}_{2,5}\left[\frac{z^2}{4}\left|^{(1-r),\nu}_{0,1/2,\nu,1,(\nu-r)}\right.\right],\nonumber
\end{eqnarray}
where $Re(\nu)>0, Re(z) > 0.$
        Next we will prove the formula [''depleted'',8,15,16,17]
\begin{eqnarray}
I_4(z,\delta,b,\nu)&\stackrel{def}{=}&\int^\infty_0 y^{\nu-1}exp[-\left\{y+by^\delta+zy^{\frac{1}{2}}\right\}]dy\label{line1}\\
&=&\sum^\infty_{r=0}\frac{(-b/z)^r}{(r)!}H^{2,0}_{0,2}\left[z\left|_{(r,1),(\nu+r\delta+r/2,1/2)}\right.\right]\nonumber\\
\end{eqnarray}                                       
 where $Re(\nu)> 0, Re(z) >0, Re(b)>0\,\,  \mbox{and}\,\, \delta>0.$
To establish the integral formula (25), we see that in view of (6), it can be written as 
\begin{equation}
I_4(z,\delta,b,\nu)=\int_0^\infty y^{\nu-1}exp(-y)\frac{1}{2\pi\omega}\int_L\Gamma(s)y^{s/2}(by^{\delta+1/2}+z)^{-s}dsdy
\end{equation} 
On employing the formula
\begin{equation}
(1+x)^{-\alpha}=\sum^\infty_{r=o}\frac{(\alpha)_r}{(r)!}(-x)^r,\,\,|x|<1,
\end{equation}
and reversing the order of integration and summation, the equation (26) transforms into the form
\begin{eqnarray}
I_4&(z,\delta,b,\nu)&=\sum^\infty_{r=0}\frac{(-b/z)^r}{(r)!}\frac{1}{2\pi\omega}\int_L\Gamma(s+r)z^{-s}\int^\infty_0 y^{\nu+(\delta+\frac{1}{2})r+\frac{s}{2}-1}e^{-y}dyds\nonumber\label{line1}\\
&=&\sum^\infty_{r=0}\frac{(-b/z)^r}{(r)!}\frac{1}{2\pi\omega}\int_L\Gamma(s+r)\Gamma[\nu+(\delta+\frac{1}{2})r+\frac{s}{2}]z^{-s}ds,\nonumber
\end{eqnarray}
which, when interpreted with the help of (10), yields the desired  result (25). 
It may be noted that the result (25) can be expressed in terms of the G-function in the form :
\begin{equation}
I_4(z,\delta,b,\nu)= \frac{1}{\pi^{1/2}}\sum^\infty_{r=0}\frac{(-2b/z)^r}{(r)!}G^{3,0}_{0,3}\left[\frac{z^2}{4}\left|_{\frac{r}{2},\frac{r+1}{2},(\nu+r\delta+\frac{r}{2})}\right.\right],
\end{equation}
where $Re(\nu) > 0, Re(b) > 0, Re(z) > 0\,\, \mbox{and}\;\;\delta>0$.
Further it is interesting to observe that, as b tends to zero, (28) reduces to (16).
\section{Relation of astrophysical thermonuclear functions to Kraetzel functions}

     It is not out of place to mention that the function represented by the integral (5) in case $z =\infty$ has been studied by Kraetzel  and is known as Kratzel function in the literature. It may be noted that Kraetzel  [18] introduced the integral transform 
\begin{equation}
(K_\nu^\rho f)(x)=\int^\infty_0 Z^\nu_\rho(xt) f(t)dt\;\; (x>0),
\end{equation}
involving the kernel function 
\begin{equation}
Z^\nu_\rho(x)= \int^\infty_0 t^{\nu-1} exp[-t^\rho-\frac{x}{t}]dt \,\,(\rho>0,\,\nu \in C).
\end{equation}
Further he  has  also investigated the asymptotic behavior of  $Z^\nu_\rho(x)$ as $x\rightarrow 0$ and $x\rightarrow \infty$  in [18].\\

Finally it may be mentioned here that the integral described here for $z = \infty$  plays not only an important role in the study of astrophysical thermonuclear functions  but also for Bessel-type fractional ordinary and partial differential equations [19].\par
\bigskip
\noindent
{\bf References}\par
\medskip
\noindent
[1] C.L. Critchfield, in: Cosmology, Fusion and Other Matters:\par 
George Gamow Memorial Volume, Ed. F. Reines, Colorado Associated\par 
University Press, Colorado, 1972, 186.\par
\noindent
[2] A.M. Mathai, H.J. Haubold, Modern Problems in Nuclear and Neutrino\par
Astrophysics, Akademie-Verlag, Berlin, 1988.\par
\noindent
[3] W.A. Fowler, Rev. Mod. Phys. 56 (1984) 149.\par
\noindent
[4] R. Davis Jr., Rev. Mod. Phys. 75 (2003) 985.\par
\noindent
[5] C. Tsallis, in: Nonextensive Entropy: Interdisciplinary Applications,\par 
Eds. M. Gell-Mann, C. Tsallis, Oxford University Press, New York,\par 
2003, 1.\par
\noindent
[6] H.J. Haubold, A.M. Mathai, SIAM Rev. 40 (1998) 995.\par
\noindent
[7] A.M. Mathai, R.K. Saxena, The H-function with Applications\par
in Statistics  and Other Disciplines, Wiley, New York, 1978.\par
\noindent
[8] W.J. Anderson, H.J. Haubold, A.M. Mathai,\par
Astrophys. Space Sci. 214 (1994) 49.\par
\noindent
[9] M. Aslam Chaudhry, S.M. Zubair, On a Class of Incomplete Gamma\par 
Functions with Applications, Chapman and Hall/CRC, Boca Raton, 2002.\par
\noindent
[10] A.M. Mathai, R.K. Saxena, Generalized Hypergeometric Functions\par
with Applications in Statistics and Physical Sciences, Lecture Notes\par 
in Mathematics, Series No. 348, Springer-Verlag, Heidelberg and\par 
New York, 1973.\par
\noindent
[11] A.M. Mathai, A Handbook of Generalized Special Functions for\par
Statistics and Physical Sciences, Oxford University Press, Oxford, 1993.\par
\noindent
[12] A. Erd\'elyi, W. Magnus, F. Oberhettinger,  F.G. Tricomi,\par
Higher Transcendental Functions, Vol. I, McGraw-Hill,\par
New York-Toronto-London, 1953.\par
\noindent
[13] H.J. Haubold, A.M. Mathai, Studies Appl. math. 75 (1986) 123.\par
\noindent
[14] S.S. Vasil'ev, G.E. Kocharov, A.A. Levkovskii, Izvestiya Akademii Nauk\par 
SSSR, Seriya Fizicheskaya 39 (1975) 310.\par
\noindent
[15] H.J. Haubold, A.M. Mathai, J. Appl. Math. Phys. (ZAMP) 37 (1986)\par 
685.\par
\noindent
[16] D.D. Clayton, E. Dwek, M.J. Newman, R.J. Talbot Jr., Astrophys. J.\par 
199 (1975) 494.\par
\noindent
[17] M. Coraddu, G. Kaniadakis, A. Lavagno, M. Lissia, G. Mezzorani,\par 
P. Quarati, Brazilian J. Phys. 29 (1999) 153.\par
\noindent
[18] E. Kraetzel, in: Generalized Functions and Operational Calculus,\par
Proc. Intern. Conf., Varna, 1975, Bulg. Acad. Sci., Sofia 1979, 148.\par
\noindent
[19] J. Rodrigues, J.J. Trujillo, M. Rivero, in: Differential Equations,\par 
Xanthi, 1987, Lecture Notes in Pure and Applied Mathematics,\par 
Series No. 118, Denver, New York, 1989, 613.\par

\end{document}